\documentclass{jetpl}
\twocolumn

  \usepackage{graphicx}
 \usepackage{bm} 

\lat

\title{Tkachenko waves}

\rtitle{Tkachenko waves}

\sodtitle{Tkachenko waves}

\author{E.\,B.\,Sonin
\/\thanks{e-mail: sonin@cc.huji.ac.il}}

\rauthor{E.\,B.\,Sonin}

\sodauthor{Sonin}

\address{
Racah Institute of Physics, Hebrew University of
Jerusalem, Jerusalem 91904, Israel}

\abstract{This is a short review of theoretical and experimental studies of Tkachenko waves starting from their theoretical prediction by Tkachenko about 50 years ago up to their unambiguous experimental observation in the Bose--Einstein condensate of cold atoms. }

\begin{document}
\bibliographystyle{plain}

\newcommand{\be}{\begin{equation}}
\newcommand{\ee}[1]{\label{#1}\end{equation}}
\newcommand{\bem}{\begin{eqnarray}}
\newcommand{\eem}[1]{\label{#1}\end{eqnarray}}
\newcommand{\eq}[1]{Eq.~(\ref{#1})}
\newcommand{\Eq}[1]{Equation~(\ref{#1})}
\newcommand{\vp}[2]{[\mathbf{#1} \times \mathbf{#2}]}

\maketitle

\section{Introduction}

In  the year 2013 Vladimir Konstantinovich Tkachenko passed away in the age of 76. This sad event urges to review the legacy of this brilliant scientist in physics.    

His active life in physics unfortunately was very short because of health problems. He published not more than about 10 papers but what papers!  Tkachenko, being nominally (and really) an experimentalist, published the papers, which Dyson \cite{Dys} called  ``{\em a tour de force} of powerful mathematics''. Tkachenko's seminal works on a vortex lattice in superfluid helium and its oscillation were written about 50 years ago but up to now they remain actual and challenging in various areas of physics, superfluid liquids, cold-atom Bose--Einstein condensates, and astrophysics among them. 

The series of Tkachenko's papers on dynamics of vortex lattices started from the paper \cite{Tka5}, in which he calculated exactly the energy of an arbitrary periodic vortex lattice and showed that the triangular lattice has the lowest energy as in the mixed state of type II superconductors. In the second paper \cite{Tka6} he found (also exactly) the spectrum of waves in the vortex lattice  for all wave vectors in the Brillouin zone. These waves are now called Tkachenko waves. Finally in his third paper \cite{Tka9} he demonstrated that in the long-wavelength limit the Tkachenko wave is nothing else as a transverse sound wave in the vortex lattice and its frequency is determined by the shear elastic modulus.

The following short review addresses the original  theory of Tkachenko waves suggested for superfluid $^4$He and its nowadays extension on Tkachenko waves in the Bose--Einstein condensate of cold atoms, and also overviews a long and controversial story of attempts to detect Tkachenko waves experimentally first in liquid $^4$He and pulsars and then  in  Bose--Einstein cold-atom condensates, which culminated in a unambiguous observation of Tkachenko waves.


       \section{Tkachenko waves from the elasticity  theory
        of a two-dimensional vortex  crystal} \label{TkachEl}

 We start not from the exact solution but from a more transparent approach  deriving the Tkachenko wave  from the elasticity theory of the vortex lattice. 

The equation   of motion in the continuous elasticity  theory for atoms in the crystal lattice is the second Newton law:
\be
 \rho {d^2 \bm u\over dt^2} =\bm f,
      \ee{}    
where $\rho$ is the mass density, $\bm u$ is the atom displacement, and the force $\bm f$ is defined as a functional derivative of the elastic energy of the crystal:       
\be
\bm f= -{\delta E\over \delta \bm u}= -   {\partial E\over \partial  \bm u}   +  \nabla_i\left(\partial E\over \partial \nabla_i \bm u\right)= \nabla_i\left(\partial E\over \partial \nabla_i \bm u\right).
   \ee{forc}       
We took into account translational invariance, which eliminates the dependence of the energy from the constant displacement $\bm u$.

Like in the elasticity theory, in vortex dynamics one can also introduce a continuous medium approximately describing an array of discrete vortex lines. This means that one carries out averaging (coarse-graining) of the equations of hydrodynamics over rather long scales of the order of intervortex distance. The approach is accurate enough as far as parameters of the medium do not vary essentially at the intervortex distance. This approach was called in Ref.~\cite{RMP} {\em macroscopic hydrodynamics}. In contrast to the elasticity theory of  atomic crystals, the equation of vortex motion connects the force on the vortex not with an acceleration but with  velocities:
\be
-\rho {2\bm  \Omega} \times (\bm  v_L
 -\bm  v) =\bm f,
      \ee{eqMot}    
where $\bm \Omega$ is the angular velocity vector, $\bm v_L=d \bm u/dt$ is the vortex velocity, and $\bm v$ is the average velocity of the liquid. We consider the $T=0$ case when the center-of-mass velocity coincides with the superfluid velocity. The angular velocity $\Omega$ determines the vortex density $n_v=2\Omega /\kappa$, where $\kappa=h/m$ is the circulation quantum and $m$ is the mass of a particle. The forces in \eq{eqMot} are forces on all vortices  piercing a unit area.
In  classical hydrodynamics the left-hand side of \eq{eqMot} is called {\em Magnus force}.  But in the theory of superfluidity and superconductivity  they usually relate the Magnus force only with the term proportional to the vortex velocity $\bm v_L$, while the term proportional to the fluid current $\rho \bm v$ is called {\em Lorentz force}.

The equations for vortex displacements must be supplemented by the continuity equation,
\be
{\partial \rho \over \partial t}+\bm \nabla \cdot ( \rho \bm v) =0,
   \ee{cont}
and  by the Euler equation, which in the rotating coordinate frame is \cite{RMP}
 \begin{equation}
 {\partial \bm  v\over \partial t}+2\bm  \Omega \times \bm  v_L
 =-\bm  \nabla \mu. 
 \label{4.1}
\end{equation}
The continuity and the Euler equations allow to determine the liquid velocity $\bm v$ and     the chemical potential $\mu$. 

The expression for the elastic force can be obtained on the phenomenological basis taking into account hexagonal symmetry of the triangular lattice.  We consider a 2D problem  in the $xy$ plane normal to the angular velocity vector $\bm \Omega$ (the axis $z$) with no dependence on $z$. The  general expression for the elastic energy density in the 2D case is \cite{LLElas}
\bem
E_ {el} ={C_{11}\over 2}(\bm \nabla \cdot \bm u)^2
\nonumber \\
+{C_{66}\over 2}\left[\left(\nabla_y u_x+\nabla_x u_y\right)^2-4\nabla_x u_x\nabla_y u_y\right].
       \eem{4.27} 
Here $C_{11}$ is the inplane compressibility modulus, and $C_{66}$ is the shear modulus. We used here Voigt's notations for elastic moduli \cite{Love} adopted in the theory of superconductivity. \Eq{4.27}  is a particular case  of a more general expression given in Refs.~\cite{RMP,Bay83}, which took into account  the $z$ dependence. From Eqs.~(\ref{forc}) and  (\ref{4.27}) one obtains an expression for the force on vortices:
\be
\bm f= (C_{11}-C_{66}) \nabla (\bm \nabla \cdot \bm u)+C_{66} \Delta  \bm u.
    \ee{}
The term proportional to the divergence $\bm \nabla \cdot \bm u$ can be neglected in the low frequency (long wavelength) limit. Then the components of the force $f_i= -\nabla_j \sigma_{ij}$ are determined by the stress tensor 
\be
  \sigma_{ij} =-C_{66}  ( \nabla_i u_j+     \nabla_j u_i  )
                                         \ee{4.67}
for purely shear deformation.  Here subscripts $i$ and $j$ take only two values $x$ and $y$ corresponding to the two axes in the
$xy$ plane. Then  the equation (\ref{eqMot}) of vortex motion becomes
\be 
{\partial \bm u\over \partial t}=\bm v_L =\bm v +{C_{66}\over 2\Omega \rho}[\hat z \times \Delta  \bm u].
    \ee{eq-u}

It is convenient to divide  the  vortex displacement field $\bm  u(\bm r)$ and the fluid velocity field $\bm v(\bm r)$  into longitudinal
and transverse parts ($\bm  u=\bm  u_\parallel+\bm  u_\perp$, $\bm  v=\bm  v_\parallel+\bm  v_\perp$), so that 
$\bm  \nabla \cdot \bm  u_\perp =\bm  \nabla \cdot \bm  v_\perp=0$ and  $\bm  \nabla \times \bm  u_\parallel     =\bm  \nabla \times \bm  v_\parallel     =0$). In an incompressible liquid $\bm v_\parallel=0$ and the liquid velocity $\bm v=\bm v_\perp$ is purely transverse.
Then \eq{4.1} after integration over  time yields
\be
\bm v=-[2\bm \Omega \times \bm u_\parallel].
     \ee{long} 
After exclusion of $\bm v$ \eq{eq-u}   yields the equations for longitudinal
and transverse displacements $\bm   u_\parallel$ and $\bm  u_\perp$:
\bem
{\partial \bm  u_\parallel \over \partial t}={C_{66}\over 2\Omega \rho}[\hat z \times \Delta  \bm u_\perp],
                                                \eem{4.63}
\bem
{\partial \bm  u_\perp \over \partial t}=-2\bm \Omega \times   \bm u_\parallel.
                                                \eem{4.64}
Excluding  the small longitudinal displacement $\bm  u_\parallel$ from equations one obtains 
an equation similar to that for the transverse sound
in the conventional elasticity theory:
\be
{\partial^2 \bm  u_\perp \over \partial t^2}=c_T^2 \Delta  \bm u_\perp,
                                                \ee{4.65}
with 
\be
c_T=\sqrt{C_{66}\over  \rho}
      \ee{}
being the velocity of the Tkachenko wave $\propto e^{i\bm k\cdot \bm r-i\omega t}$ with the sound-like spectrum $\omega =c_T k$.         Vortices in the Tkachenko wave move on elliptical
paths, but the longitudinal component $ \bm u_\parallel$  parallel to the wave vector $\bm  k$ is proportional to a small factor $\omega/\Omega$ [see \eq{4.64}]. Thus it is fairly accurate to
consider the Tkachenko wave to be a transverse sound
wave in the two-dimensional lattice of rectilinear vortices \cite{Tka9}. Comparing Eqs.~(\ref{long}) and (\ref{4.64}) one can see that in our approximation  the liquid and the vortices move with the same velocity.
The phenomenological approach cannot provide the value of the shear modulus. But it is clear that the elastic energy is in fact the kinetic energy of the velocity field induced by the vortices, and scaling estimations show that the shear modulus should be on the order of $C_{66} \sim \rho \kappa \Omega$. Its exact value can be obtained from the exact value  of the energy of the vortex lattice obtained by Tkachenko (Sec.~3). 

All experiments on Tkachenko waves dealt with   finite cylindric liquid samples, and it is necessary to know the boundary conditions for Tkachenko cylindric waves.  We restrict ourselves with
axisymmetric modes.

The field of transverse displacements $\bm u_\perp$ may be determined by a vector potential $\bm  \Psi =\Psi \hat z$:
\begin{equation}
\bm   u_\perp =\bm  \nabla \times \bm  \Psi =-\hat z \times \bm  \nabla \Psi.
\label{5.11}
\end{equation}
The potential  $\Psi $ must satisfy the wave equation
\begin{equation}
{\partial^2 \Psi \over \partial t^2}-c_T^2 \Delta \Psi=0.
\label{5.12}
\end{equation}
Axisymmetric modes with the sound-like spectrum $\omega =c_T k$ correspond to a cylindrical wave
\bem
\Psi= \Psi_0 J_0(kr)e^{-i\omega t},
\nonumber \\
u_r\approx 0,~~u_\varphi=-{\partial \Psi\over \partial r}=k \Psi_0 J_1(kr)e^{-i\omega t},
        \eem{5.13}
where subscripts $r$ and $\varphi$ denote radial and azimuthal components in the cylindric coordinate frame   $(r,\varphi)$.

        Suppose that no external force acts upon the liquid, which fills a cylinder of the radius $R$.
Then eigenfrequencies are defined by the condition that
the total angular momentum $M$ does not vary. Since in the
Tkachenko wave the fluid and vortices move together
\bem
M=2\pi \rho \int_0^R v_\varphi r^2dr=-2i\omega \pi \rho \int_0^R u_\varphi r^2dr
\nonumber \\
=-2i\omega \pi \rho\Psi_0 R^2 J_2(kR)e^{-i\omega t}.
        \eem{5.14}
 The condition $M=0$ yields eigenfrequencies
\begin{equation}
\omega_i=j_{2,i} {c_T \over R}, 
\label{5.15}
\end{equation}
where $j_{2,i}$ denotes the $i$th zero of the Bessel function $J_2(z)$.
For the fundamental frequency $j_{2,1} = 5.14$. This is a result obtained by Ruderman \cite{Rud} who discussed Tkachenko modes in pulsars (see Sec.~4). 

The condition   $M=0$  is equivalent to the boundary condition that the azimuthal component of the momentum flux through the liquid boundary $r=R$ vanishes. This momentum flux is given by the relevant stress tensor component $\sigma_{\varphi r}$ in  cylindrical coordinates:
\be
\sigma_{\varphi r}(r)=-\rho c_T^2\left({\partial u_\varphi(r) \over \partial r}-{u_\varphi(r) \over r}\right).
    \ee{5.19}
The condition $\sigma_{\varphi r} (R)=0$ requires that 
\be 
{\partial u_\varphi(R) \over \partial r}-{u_\varphi(R) \over R}=0.
   \ee{BC1}
This   yields the same spectrum \eq{5.15} as the condition $M=0$.

\section{Exact solution of Tkachenko} \label{exact}

\begin{figure}
  \begin{center}
    \leavevmode
    \includegraphics[width=\linewidth]{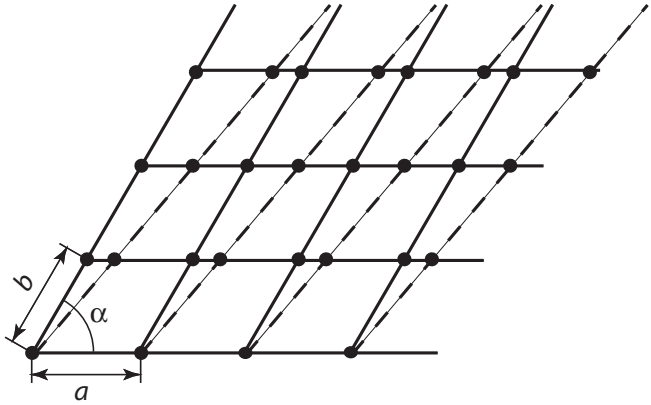}
    \caption{Fig. 1. Vortex lattice before (solid lines) and after (dashed lines) shear deformation.}
  \label{fig1lat}
  \end{center}
  \end{figure}

Tkachenko has found an exact solution for the vortex lattice and its oscillation using  the theory of elliptic functions on the complex plane \cite{Tka5,Tka6}. It is well known that a two-dimensional vector $\bm r(x,y)$ can be presented as a complex variable $z=x+iy$. Then the velocity field $v(z)=v_x+iv_y$ induced by vortices located in nodes of a vortex lattice with position vectors $z_{kl}=2k \omega_1+ 2l \omega_2$ ($k$ and $l$ are  arbitrary integers)  is given by 
\be
v(z)={\kappa \over 2\pi}[\zeta^*(z) -\lambda z^*],
    \ee{} 
 where $\lambda$ is a constant, which will be defined below, and
\be
\zeta(z) ={1\over z}  +{\sum_{k,l}}'\left( {1\over z-z_{kl}}+{1\over z_{kl}}+{z\over z_{kl}^2}\right)
     \ee{}   
is the quasiperiodic Weierstrass zeta function \cite{AS}  with two complex semi-periods $\omega_1$ and $\omega_2$ and a prime means exclusion of the term $k=l=0$ from the sum.  The quasiperiodicity conditions are 
\bem
\zeta(z+2k \omega_1) =\zeta(z)+2k \omega_1,
\nonumber \\
\zeta(z+ 2l \omega_2) =\zeta(z)+2l \omega_2.
       \eem{}
The lattice is shown in Fig.~\ref{fig1lat} for the semi-periods $ \omega_1= a/2$ and $ \omega_2= be^{i\alpha}/2$.  The unit cell area of the lattice is 
\be
A=4 \mbox {Im} (\omega_1^* \omega_2)=ab \sin \alpha.
       \ee{}
Tkachenko has shown that a lattice with arbitrary  semi-periods rotates as a solid body with the angular velocity $\Omega=\kappa/2 A$ , if one chooses $\lambda$ satisfying the condition
\be
\zeta(\omega_1) +\lambda \omega_1 =\Omega \omega_1^*.
      \ee{}
Taking into account the exact relation for  Weierstrass zeta function,
\be
\omega_2\zeta(\omega_1)- \omega_1\zeta(\omega_2)  ={i\pi\over 2},
     \ee{}   
another condition necessary for solid-body rotation is also satisfied:
\be
\zeta(\omega_2) +\lambda \omega_2 =\Omega \omega_2^*.
      \ee{}
For symmetric triangular and quadratic lattices $\lambda=0$.

The velocity field being known Tkachenko \cite{Tka5} found after ingenious manipulations with integrals over elliptic functions the exact value of kinetic energy per unit area in the rotating coordinate frame for an arbitrary vortex lattice: 
\bem
E={\rho \kappa \Omega  \over 2\pi }\left[
\ln{2|\omega_1 \omega_2|^{1/2}\over \pi r_c} -{\ln 2\over 3} \left| \theta_1'(0,\tau) \theta_1 '\left(0,-{1\over \tau}\right)\right|\right]
\nonumber \\
={\rho \kappa \Omega \over 2\pi }\left[ \ln{\sqrt{ A|\tau|}\over \pi r_c\sqrt{ \tau_I }}   -{\ln 2\over 3} \left| \theta_1'(0,\tau) \theta_1 '\left(0,-{1\over \tau}\right)\right|\right],
\nonumber \\
     \eem{enTk}
where the complex parameter 
\begin{equation}
\tau =\tau_R+i \tau_I=\omega_2 /\omega_1={b\over a}e^{i\alpha}
    \label{tau}   \end{equation} 
determines the type of the lattice,
\begin{eqnarray}
\theta_1(z, q )=-i\sum_{n=-\infty}^\infty (-1)^n q^{(n+1/2)^2} e^{i(2n+1)z} 
         \end{eqnarray} 
is one from the elliptic theta functions, and  $\theta_1'(z, q )$ is its derivative with respect to the first argument $z$. The energy has a minimum at $\tau = e^{i\pi /3}$ ($a=b=\sqrt{\kappa/\sqrt{3} \Omega}$, $\alpha=\pi/3$), which corresponds to the triangular lattice with the energy density
\[
E_m ={\rho \kappa \Omega \over 2\pi }\left( \ln{\sqrt{ A}\over r_c} - 1.321 \right).  \]

In order to find the shear modulus let us deform the triangular lattice without varying the vortex density as shown  in Fig.~\ref{fig1lat}.  
Then only the real part of $\tau$ varies proportionally to the shear deformation $u_{xy}={1\over 2}(\nabla_y u_x+\nabla_x u_y)$: $\delta \tau=\delta \tau_R = 2 u_{xy} \sin
\alpha$.  
Expanding the energy density Eq. (\ref{enTk}) with respect to $\tau_R$ and comparing it with the elastic energy  (\ref{4.27}) with $\bm \nabla \cdot \bm u=0$  one obtains the exact value of the shear modulus:
\be
C_{66}=\rho c_T^2={\rho \kappa\Omega\over 8\pi}.
                                                \ee{4.34}.

\section{Tkachenko waves in pulsars}\label{puls}

       Some features of pulsar behavior have been explained
by the hypothesis that the rotating inner matter of pulsars is
in the superfluid state and is threaded by vortex lines. 
Among  such features were sudden spin-ups
of pulsars (glitches)  and slow  relaxation after the glitch \cite{Dys,Alp8,And8}. In addition, very slow oscillations of the
Crab pulsar's period have been observed \cite{Dys}. Ruderman \cite{Rud} has associated this remarkable phenomenon with
Tkachenko waves. He considered waves in a cylinder, ignoring the difference between cylindrical and
spherical geometry. Inserting into \eq{5.15} the data for
the pulsar in the Crab nebula ($\Omega=200$ rad/sec, $ R=10^6$ cm,
$\kappa=2\times 10$ cm$^2$/sec, Ruderman found that the oscillation period for the fundamental mode $s=1$
should be
\[ T={2\pi \over \omega_R}=9.73\times 10^6~\mbox{sec}=3.75~\mbox{months}
\]
in good agreement with the observed period of $\sim 3$
months.   Dyson \cite{Dys} argued that it is difficult to think of any other internal motion, which would have a time scale as long as this.

Ruderman's model was rather
idealized even for very long cylinders when pinning of
vortices to the solid surface is important. In
pulsars the solid crust confining the neutron matter plays
the role of a solid surface. More on this issue was discussed in Ref.~\cite{RMP}.

Later the interest to interpretation of pulsar oscillations in the terms of the Tkachenko mode  declined to some extent and other interpretations were suggested. But recently some publications urged to return to the Tkachenko-mode interpretation of long-period pulsar oscillations \cite{Sedr,Ppv,Sha}.

 \section{ Search of Tkachenko waves 
       in superfluid $^4$He}\label{TkachE}\label{VI.E}
        
        The first attempt to observe a Tkachenko wave in a laboratory   was
undertaken by Tkachenko himself in the 1970s in a study of
torsion oscillations of a light cylinder immersed into  rotating superfluid  $^4$He and suspended by a thin
fiber \cite{Tka74}. The oscillating cylinder cannot drag the superfluid component of the liquid but it does drag the normal one. The latter makes superfluid vortices to oscillate via mutual friction. No conclusive data were obtained in this experiment. A later analysis of this case (see Sec.~VIII.D in Ref.~\cite{RMP}) showed that an essential contribution of the Tkachenko mode  would be possible for rather fast rotation inaccessible at that time.

        Further efforts to discover Tkachenko waves experimentally
were stimulated by Ruderman's theory explaining long-period
oscillations of the pulsar rotation velocity. For simulation of the process in pulsars,
J. Tsakadze and S. Tsakadze \cite{Tsa73,Tsa75}  studied free rotation of buckets of various shapes,
cylindrical included, filled with He II,
and revealed rotation-period oscillations superimposed on
the steady deceleration of rotation. The oscillations
disappeared above the $\lambda$ point that proved their
superfluid nature. But the oscillation frequencies observed
for cylindrical vessels were nearly eight times
higher than the fundamental frequency predicted by Ruderman \cite{Rud} for this geometry.
This disagreement was explained by three-dimensional effects of pinning and bending of vortices \cite{Son76,RMP}.  These  effects transform the Tkachenko mode into a mixed mode combining the Tkachenko wave and the inertial wave with the spectrum \cite{Son76,Wil}
\be
\omega=\sqrt{4\Omega^2 {p^2\over k^2+p^2}+c_T^2 k^2}.
                                                \ee{4.55}
Here $p$ is the $z$ component of the wave vector, which appears in the mixed plane wave $\propto e^{i\bm k\cdot \bm r+ip z-i\omega t}$ as a result of pinning at surfaces normal to the rotation axis (the axis $z$). Without the quantum contribution $c_T^2 k^2$ this is an inertial wave well known in hydrodynamics of rotating classical fluids \cite{Gri}.  The quantum contribution depends on the container radius $R$ since for the Tkachenko-wave resonance $k \sim 1/R$. The Tkachenko velocity usually is very small (of order 1 cm/sec). As a result, the frequency $\omega=c_Tk$ of the pure Tkachenko mode is much smaller than $\Omega$. Then according to \eq{4.55} even rather weak pinning leading to rather weak vortex bending (small $p$) can  strongly influence the mode frequency. As a result, the inertial-wave contribution essentially exceeds the quantum (Tkachenko) contribution. The inertial-wave contribution grows with decreasing of the height $L$ of helium in the container (the length of vortices). 
J. Tsakadze and S. Tsakadze \cite{Tsa73,Tsa75} used cylindric containers of moderate aspect ratio $L/R$ when the quantum Tkachenko contribution was negligible.  Therefore they observed  the inertial-wave resonance. This was proven by experimental detection \cite{Tsa76} of properties
predicted by the theory of the initial-wave resonance \cite{Son76}.    The observed oscillation
frequency depended on $L$ and on the smoothness of the bottom,
but did not depend on the container radius $R$ (see more detailed comparison and discussion in Refs.~\cite{RMP,Tsa76,Tsa80}).

        In further experiments, S. Tsakadze \cite{Tsa78} used longer
cylindrical containers in an effort to reach  the conditions
when pure Tkachenko waves are possible. He could not
do it completely, because it required impractical containers
with too large  ratios $L/R$, but he managed to come fairly close  to the case when 
the Tkachenko contribution to the oscillation frequency was of 
the same order as the inertial-wave  contribution. S. Tsakadze noticed an essential deviation from the frequency of the inertial wave resonance. The deviation roughly agreed with what was expected from the mixed-wave resonance when the classical and the quantum contributions to the spectrum \eq{4.55} were of the same order.
This was the first experimental evidence  of the Tkachenko elasticity and consequently of crystalline order in the vortex lattice.

The next attempt of observation of the Tkachenko wave was undertaken by Andereck {et al.} \cite{And0,And2}, who claimed that they saw Tkachenko waves in the experiment on torsional oscillations of a  pile-of-disks immersed  
into a rotating superfluid $^4$He. They observed a resonance, which they connected with a peak in the density of state caused by a minimum of the spectrum \eq{4.55}  at given $p$. But the theoretical analysis of  Andereck et al.  left unresolved a serious problem (by admission of the authors themselves; see p.288 in the paper by Andereck {et al.} \cite{And2}): how can the oscillations of disks, introducing perturbations with wavelengths of the order of the disk radius, generate waves  whose wavelengths are an order of magnitude smaller than the radius of the
disks? Andereck {et al.} believed that they observed Tkachenko modes for very low aspect ratio $L/R$ ($L$ in their case was a small distance between disks), which was in conflict with the conclusion that because of pinning observation of Tkachenko modes requires very high aspect ratio. Later it was demonstrated \cite{Son83} (see also Ref.~\cite{RMP}) that the resonance observed by  Andereck {et al.} can be readily explained as a predominantly inertial-wave resonance  without any contribution of Tkachenko rigidity.

In summary, experimental observation of Tkachenko waves in superfluid $^4$He encountered serious problems connected with pinning of vortices at solid surfaces containing superfluids. Evidences of the Tkachenko mode  were rather circumstantial  and did not allow a decisive quantitative comparison with the theory. A breakthrough in experimental observation of Tkachenko waves became possible after discovery of a new type of superfluids:
Bose--Einstein  condensates of cold atoms. In these new superfluids a superfluid sample being confined by a potential trap has no contacts with any solid surface. This  excludes the main hurdles for observation of pure Tkachenko waves: pinning and competition  with the inertial-wave resonance. 
However, a number of assumptions used in the Tkachenko theory became invalid: incompressibility and homogeneity  of the liquid. This required revision of the theory, which will be discussed in the following sections.

   \section{Tkachenko wave in a compressible perfect fluid}  \label{Tk-b}
        
        For a discussion of the effect of compressibility on
vortex oscillations we need to return to the general linear equations of motion, Eqs.~(\ref{cont}), (\ref{4.1}), and (\ref{eq-u}). Let us neglect first inhomogeneity of the liquid. Then the equations of motion have 
plane wave solution $\propto e^{i\bm k\cdot \bm r -i\omega t}$ and  after linearization become
\be
-i\omega \rho' +\rho \bm k \cdot \bm v=0,
                                                \ee{4.68}
\be
-i\omega \bm v+2\bm \Omega \times \bm v_L +i \bm k  \mu'=0,
                                                \ee{4.69a}
\be 
-i\omega  \bm u =\bm v -{C_{66}k^2\over 2\Omega \rho}[\hat z \times \bm u],
    \ee{eq-uF}
where $\rho'$ and $\mu'$ are small corrections to the mass density $\rho$ and the chemical potential $\mu$ induced by the propagating wave. Using the  relation $\mu'= (c_s^2/\rho)\rho'$ where $c_s$ is the sound velocity one can exclude the density correction $\rho'$ and obtain the equation connecting the velocities $\bm v$ and $\bm v_L$:
\be
-i\omega \bm v+2\bm \Omega \times \bm v_L +{c_s^2\over \rho }i \bm k {\bm k \cdot \bm v\over i\omega} =0.
                                                \ee{4.69}
We should solve two 2D vector equations (\ref{eq-uF})  and  (\ref{4.69}). One can divide the velocity and the displacement 
fields into longitudinal (parallel to $\bm k$) and transverse (perpendicular to $\bm k$) parts again. We consider only  low frequencies $\omega \ll c_sk$ excluding usual sound waves. Then the transverse velocity $-i\omega \bm u_\perp$ approximately coincides with the transverse velocity $\bm v_\perp$ of the liquid and excluding small longitudinal velocity $-i\omega \bm u_\parallel$ Eqs.~(\ref{eq-uF})  and  (\ref{4.69}) reduce to two equations for the longitudinal and the transverse liquid velocities $\bm v_\parallel$ and $\bm v_\perp$:
\bem
2\Omega i\omega v_\parallel =-\omega^2 v_\perp + {C_{66} k^2\over \rho}v_\perp, 
\nonumber \\
2\Omega i\omega v_\perp=- c_s^2 k^2  v_\parallel.
   \eem{plCom}
This yields the dispersion relation \cite{Son76} 
\be
\omega^2=\frac{c_s^2c_T^2 k^4}{c_s^2k^2+4\Omega^2}.
                                                \ee{4.73}
This dispersion relation also follows from a more general expression obtained by Volovik and Dotsenko \cite{Vol80} using the method of Poisson brackets.
In  the limit $c_s\to \infty$ \eq{4.73} yilds the spectrum of the Tkachenko wave in an incompressible liquid. The  compressibility strongly alters the spectrum of
this wave at small $k \ll 2\Omega /c_s$, making it parabolic:
\be
\omega=\frac{c_sc_T}{2\Omega}k^2.
                                                \ee{4.74}

In superfluid $^4$He and $^3$He the effect of compressibility on inertial waves 
is rather academic, because the space scale $c_s/\Omega$
at which the incompressible-fluid hydrodynamics becomes invalid is extremely large (of order hundreds
of meters) and is not relevant to any real laboratory experiment. So at that time, when this effect was first analyzed  \cite{Son76,RMP}, it was considered  as a theoretical curiosity, or belonging to some astrophysical
applications. The situation became essentially different  after discovery of  the BEC of cold atoms.
In contrast to   the both helium superfluids, BEC is a weakly interacting Bose gas with very low sound speed and very high compressibility. Importance of
high liquid compressibility for Tkachenko waves in BEC was pointed out by Baym \cite{Baym}.

In a compressible liquid  centrifugal forces make the liquid density essentially inhomogeneous at the scale $c_s/\Omega$. At the distance $r\sim c_s/\Omega$ from the rotation axis the linear velocity $\Omega r$ of solid body rotation becomes of the same order as the sound velocity $c_s$.  Therefore, the analysis of the homogeneous liquid presented above is purely illustrative and cannot be directly applied to practical problems arising in experiments of Tkachenko waves in BEC of cold atoms. One should take into account inhomogeneity of the fluid.

Let us now consider axisymmetric Tkachenko modes in a rotating finite BEC cloud of pancake geometry. One can ignore variations along the axis of the pancake, and the problem becomes two-dimensional.  The 2D cloud is trapped by the parabolic potential ${1\over 2}m\omega_\perp^2 r^2$.
The Thomas--Fermi approximation \cite{PS,Ued} yields an inverted parabola distribution of the mass density:
$\rho(r) =\rho(0)(1-r^2/R^2)$. Here
$R=\sqrt{2}c_s(0)/\sqrt{\omega_\perp^2-\Omega^2}$ is the cloud radius (Thomas--Fermi radius) and $c_s(0)$ is the
sound velocity at the symmetry axis $r=0$. The radius $R$ grows with the angular velocity because of the effect of centrifugal forces. The Tkachenko mode in such a geometry was investigated numerically with solving the equations of Gross--Pitaevskii theory  \cite{Jap,Bak}. On the other hand, in experiments the cloud size $R$ essentially exceeded the intervortex distance. Therefore a simpler approach based on the macroscopic hydrodynamics  explained in Sec. 2 
 can provide a deeper insight into physics of the phenomenon \cite{ST}. 

Since compressibility effect becomes important at $k\sim \Omega/c_s$ and the eigenvalues of $k$ are expected to be of the order of $1/R$ the compressibility effect is essential if the parameter 
\begin{equation}
 s ={\Omega R \over \sqrt{2} c_s(0) }={\Omega  \over
\sqrt{\omega_\perp^2-\Omega^2}}
  \label{comp} \end{equation} 
is of order of unity or more. Thus at rapid rotation of the BEC with angular velocity $\Omega$ close to the trap frequency $\omega_\perp$ liquid compressibility should be taken into account. 

 The equations of motion (\ref{plCom})  for plane waves in a homogenous compressible liquid can be transformed to those describing  a monochromatic axisymmetric cylindric mode $\propto e^{-i\omega t}$  in the cylindric system of coordinates:
\begin{equation}
2 \Omega i\omega v_r=-\omega ^2 v_t-{1\over \rho(r)r^2} {\partial \over \partial
r}\left[\rho(r)c_T^2 r^3{\partial \over \partial r}\left(v_t\over r\right)\right], 
        \label{eq1}\end{equation}
\begin{equation}
 2 \Omega i\omega v_t={\partial \over \partial r}\left[ {c_s^2(r)\over \rho(r)r}{\partial
(\rho(r) rv_r)\over \partial r}\right] .
    \label{eq2}\end{equation}
 Longitudinal and  transverse components correspond to  radial (subscript $r$) and  azimuthal (subscript  $\varphi$) components respectively. Now $\rho$, $c_s$ and $C_{66}=\rho c_T^2 $ depend on the distance $r$ from the rotation axis, but the Tkachenko velocity $c_T$ does not depend on density. For a weakly interacting Bose gas $c_s^2$ is proportional  to the density $\rho$. Therefore
the ratio $c_s^2/\rho$ is a constant equal to its value $c_s^2(0)/\rho(0)$ in the cloud center $r=0$. 

As well as in an incompressible liquid, the flux of the azimuthal component of the momentum through the liquid boundary $r=R$ given by \eq{5.19} must  vanish. Since the stress tensor (momentum flux) is proportional to
$\rho$ and the latter vanishes at $r=R$, it looks that the momentum flux through the boundary vanishes independently from
whether the boundary condition  \eq{BC1} is satisfied or not. But this is not true.  Solving the equation of motion (\ref{eq1}) close to $r=R$ by expansion in small $(R-r)/R$ one obtains that  $v_t \approx
r[C_1 +C_2 \ln (R-r)]$, where $C_1$ and $C_2$ are arbitrary constants. The component $\propto C_2$ diverges at $r\to R$ and gives a finite contribution to the stress tensor despite the factor $\rho
\propto R-r$. So this component should be
absent. This requirement is satisfied only if the boundary condition Eq. (\ref{BC1}) takes place. 

In addition to the boundary condition (\ref{BC1}) we need the second boundary condition imposed on radial
liquid velocity $v_r$. We use the arguments similar to those used for derivation of Eq. (\ref{BC1}). The
total mass balance requires that the radial mass current
$\rho(r) v_r(r)$ at the BEC cloud  border  $r=R$ vanishes.  Solving Eq. (\ref{eq2}) at $r \approx R$ by series expansion [again neglecting terms $\sim
(R-r)^2$] one obtains:
\begin{equation}
 v_r(r)= { \Omega i\omega v_t(R)R\over c_s(0)^2}{R-r\over 2}+C_1 \left(1+{R-r
\over R}\right)+{C_2\over R-r}.
   \end{equation}
The divergent component $\propto C_2$  gives a finite mass flow at the border and should be eliminated. 
Taking a
derivative from $v_r$  and excluding the constant $C_1$ from the expressions for $v_r$ and its derivative we receive the boundary condition imposed on $v_r$:
\begin{equation}
{dv_r(R)\over d r}+{v_r(R)\over R}= -{i\omega\Omega R\over 2 c_s(0)^2}v_t(R)~.
  \label{BC-2} \end{equation}

It is useful to introduce the dimensionless variables for Eqs.~(\ref{eq1}) and (\ref{eq2}):
\be 
\tilde r={r\over R},~~\tilde \omega ={\omega R\over c_T},~~\tilde v_r={iv_r\over c_T},~~\tilde v_t={v_t\over c_s(0)}.
    \ee{}
Then Eqs.~(\ref{eq1}) and (\ref{eq2}) and the boundary conditions to them become purely real and depend only on the compressibility parameter $s$  given by \eq{comp}. Solving them numerically one obtains reduced eigenfrequencies $\tilde \omega_i=f_i(s)$ as functions of $s$.  At large $s$ the eigenfrequencies $\tilde \omega_i=\gamma_i /s$ are  inversely proportional to $s$. The first two eigenfrequencies correspond to $\gamma_1=9.66$ and $\gamma_2=22.8$. Returning back to dimensional frequencies at rapid rotation ($\omega_\perp -\Omega \ll \omega_\perp$) their values are
\be
\omega_i={\gamma_i\over s} {c_T\over R} \approx  {\sqrt{2} \gamma_i } {c_T\over c_s(0)}(\omega_\perp-\Omega).
    \ee{omegRap}
Qualitatively this simple expression (aside from a numerical factor) follows from the dispersion relation (\ref{4.74}) for Tkachenko plane waves taking into account that the eigenmodes of the cloud correspond to wave numbers $k\sim 1/R$.

\begin{figure}
  \begin{center}
    \leavevmode
    \includegraphics[width=\linewidth]{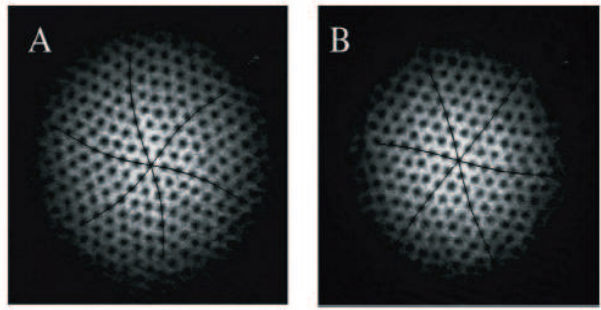}
 \caption{Fig. 2. Tkachenko mode excited in a rotating Bose--Einstein condensate of $^{87}$Rb atoms by a pulse in two moments after the pulse \cite{Cod}.  Line are sin fits to distortions of the vortex lattice by the Tkachenko mode.}
  \label{figBould}
  \end{center}
  \end{figure}

\begin{figure}
  \begin{center}
    \leavevmode
    \includegraphics[width=\linewidth]{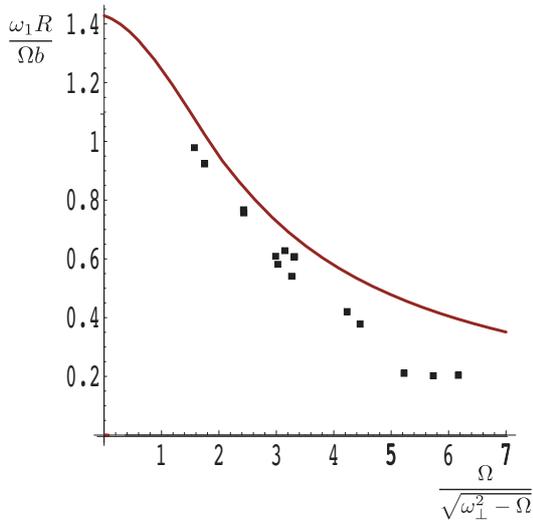}
    \caption{Fig. 3. Comparison between the theory (solid line) and the experiment (black squares). Here $\omega_1$ is the first Tkachenko eigenfrequency and $b=\sqrt{\kappa/\sqrt{3} \Omega}$ is the intervortex distance.}
  \label{fig1}
  \end{center}
  \end{figure}

Let us now address the experiment, which provided the first unambiguous experimental observation of  Tkachenko waves. It  is remarkable that in Bose--Einstein condensates of cold atoms  it was possible to observe Tkachenko waves {\em visually}. Figure~\ref{figBould} shows the image of Tkachenko wave obtained by Coddington {\em et al.} \cite{Cod} in a rotating Bose--Einstein condensate of $^{87}$Rb atoms. 
In Fig.~\ref{fig1} black squares show experimental points
 \cite{Cod} plotted in our dimensionless variables by I. Coddington. They were obtained for various parameters, but
collapse on the same curve, as expected from the present analysis. The solid line in the same figure shows 
the numerically found first
eigenfrequency $
\omega_1$  plotted as a function of
$\Omega/\sqrt{\omega_\perp^2-\Omega^2}$ (solid line).  Quantitative agreement between
the theory and the experiment looks quite good. Coddington {\em al.} \cite{Cod} measured also the 
ratio of the two first frequencies $\omega_2/\omega_1 =1.8$  at $\Omega/\omega_\perp=0.95$, which
corresponds to $s=3.04$, The present theory predicts the ratio $\omega_2/\omega_1 =2.09$.

The agreement becomes worse at larger
$s=\Omega/\sqrt{\omega_\perp^2-\Omega^2}$. This can be connected with violation of the assumption, which the theory was based upon: The vortex lattice is an array of singular vortex lines with the core size (of the order of the coherence length $\xi \sim \kappa/c_s$)  much less than the intervortex distance $b$. 
One can call this the Vortex Line Lattice (VLL) regime. With $\Omega $ approaching  the trap frequency $\omega_\perp$ the cloud radius $R$ grows, and if the total number of particles remain constant the particle density decreases. The sound velocity $c_s$ decreases also and the core radius grows.  When $\xi$ becomes of the same order as the intervortex distance $b$ the vortex cores start to overlap like in the mixed state of a type-II
superconductor close to the second critical magnetic field $H_{c2} \sim \Phi_0/\xi^2$
($\Phi_0$ is the magnetic flux quantum).  At the critical magnetic field $H_{c2}$  the phase  transition to the normal state takes
place. However, in a rotating BEC there is no phase transition at the
``critical'' angular velocity $\Omega_{c2}
\sim \kappa/\xi^2$. Instead the crossover to a new regime takes place: At $\Omega \gg
\Omega_{c2}$ all atoms condense into a state, which is a coherent superposition of
single-particle states  in the Lowest Landau Level (LLL) similar to that of a charged particle in a
magnetic field (the LLL regime).  This interesting regime, which is called also the mean-field quantum Hall regime,  is now the subject of intensive experimental and theoretical investigations
 \cite{Ued}.    

\section{Search of Tkachenko waves in the LLL regime}

The plausible approach to the vortex dynamics in the LLL regime is that the phenomenological theory developed in Sec.~6 
is still valid in this regime  but one should revaluate elastic moduli and the expressions for the sound and the Tkachenko velocities $c_s$ and $c_T$. For this revaluation we consider an infinite periodic vortex lattice in an infinite uniform liquid, neglecting
first the trapping potential but taking into account interaction.   In the Gross-Pitaevskii theory the Gibbs thermodynamic potential is
\begin{eqnarray} 
G=-m \mu |\psi|^2+{\hbar^2\over 2m}\left|\left(-i \nabla -{2\pi\over \kappa}
\bm  v_0
\right)\psi\right|^2 +{g\over 2} |\psi|^4 .
         \label{G} \end{eqnarray} 
Here $\psi$ is the BEC wave function, $\mu$ is the chemical
potential, $g$ is the interaction constant, and $\bm  v_0=[\bm  \Omega \times \bm  r]$ is
the velocity of the solid body rotation. Let us consider the
gauge transformation $\psi \rightarrow \psi e^{i\phi}$, $\bm  v_0 \rightarrow \bm  v_0
+(\kappa/2\pi)\bm  \nabla \phi$ with constant $\bm  \nabla \phi$.
 The  Gibbs potential Eq. (\ref{G}) is invariant with respect to this gauge transformation 
if it is accompanied by translation, which corresponds to a shift of
the rotation axis.

The exact wave function
for this state was found in the classical work by Abrikisov \cite{Abr} for type II
superconductors close to
$H_{c2}$, and later it was generalized for an arbitrary unit cell of the vortex lattice
 \cite{Kl,SG}.  As well as for type-II superconductors close to $H_{c2}$, in zero-order approximation one
can neglect interaction (nonlinear term $\propto |\psi|^4$). Then the linear
Schr\"odinger equation is similar to that for a charged particle in a uniform magnetic field: 
\begin{equation}
m\mu\psi=-{\hbar^2\over 2 m} \left[\left({\partial \over
\partial x} -i{2\pi v_{0x}\over
\kappa} \right)^2 + \left({\partial \over
\partial y} -i{2\pi v_{0y}\over
\kappa} \right)^2\right]\psi~.
   \label{Sch}     \end{equation} 
   At $\mu=\hbar \Omega/m$ it has a solution, which
corresponds to the lowest Landau level:
\begin{equation}
\psi_k \propto \exp\left[ikx-{(y-y_k)^2\over 2l^2}\right]~,
    \label{psik}    \end{equation} where $l^2=\kappa/4\pi\Omega$ and $y_k=-l^2k$. The
solution is given for the gauge with $\bm  v_0(-2\Omega y, 0)$. The frequency $2\Omega$ is
the analog of the cyclotron frequency
$\omega _c =eH/mc$ for an electron in a magnetic field. If we consider a square $L\times L$
with periodic boundary conditions, then $k=-2\pi n /L$ with the integer $n$. Using the
condition  $0<y_k<L$, one can see that the integer $n$ should vary from zero to the integer
closest to $L^2 /2\pi l^2$. This is the total number of LLL states, which is exactly equal to the number of vortices $2\Omega L^2 /\kappa$. All these states are  orthogonal to each other and have the same
energy. The degeneracy is lifted by taking into account the interaction energy. The
solution, which corresponds to the periodic vortex lattice with one quantum per lattice
unit cell,  is \cite{SG}
\begin{equation}
\psi_=\sum_n C_n\exp\left[inkx-{(y+l^2nk)^2\over 2l^2}\right]~,
        \end{equation} where $C_{n+1}=C_n \exp(2\pi ib \cos \alpha/a )$, $a$, $b$, and the
angle $\alpha$ are the parameters of the unit lattice cell (see Fig.  \ref{fig1lat}). 

This solution yields the thermodynamic potential of the infinite BEC in the LLL regime averaged over the vortex lattice unit cell:
\begin{equation} G=(-m\mu +{\hbar\Omega}) n+{g\over 2} \beta n^2~,
      \label{GL} \end{equation} 
where $n=\langle |\psi|^2\rangle$ is the average
particle density and the parameter  \cite{SG}
\be
\beta={\langle |\psi|^4\rangle\over \langle
|\psi|^2\rangle^2}=\sqrt{\tau_I}\left\{\left|\theta_3(0, e^{2\pi i \tau })\right|^2
+\left|\theta_2(0, e^{2\pi i \tau })\right|^2\right\}
    \ee{beta}  
    depends on lattice parameters $a$, $b$,
and $\alpha$ via the complex parameter $\tau$ determined by \eq{tau}.
Here 
\begin{eqnarray}
\theta_2(z, q )=\sum_{n=-\infty}^\infty q^{(n+1/2)^2} e^{i(2n+1)z}, \nonumber \\
\theta_3(z, q )=\sum_{n=-\infty}^\infty  q^{n^2} e^{i 2nz}  
       \end{eqnarray} 
are theta functions \cite{AS}.  The minimum of the interaction energy
corresponds to the triangular vortex lattice with $\beta=1.1596$, 
$a=b=2l\sqrt{\pi/\sqrt{3}}$, $\alpha=\pi/3$.  According to Eq. (\ref{GL}) the Gibbs
potential has a minimum  at the particle density $n=(m\mu -\hbar \Omega) / \beta g$.  This allows to determine the sound velocity.
\be
c_s=\sqrt{\rho{\partial \mu \over \partial \rho}} = \sqrt {\beta gn \over m}.
         \ee{}
This insignificantly differs from the expression for the sound velocity in the VLL regime by the factor $\sqrt{\beta}$, which is very close to unity.

Calculation of the shear elastic modulus $C_{66}$ in the LLL regime is similar to that in the VLL regime. 
Deforming  the
triangular lattice as shown  in Fig.~\ref{fig1lat}, the real part $\tau_R$ of the complex parameter
$\tau$ varies proportionally to the shear deformation $ u_{xy}$.  Expanding the expression Eq. (\ref{beta}) for $\beta$ and comparing  the term
$\propto \delta \tau_R^2$ in the thermodynamic potential, Eq. (\ref{GL}), with the elastic
energy Eq.~(\ref{4.27}), one obtains the value of the shear modulus:
\begin{equation} 
C_{66}={gn^2\over 2} {\partial^2 \beta \over \partial \rho^2}\sin^2
\alpha= 
0.2054 \rho c_s^2.
    \label{C-2}   \end{equation}
This agrees with the value of the shear modulus
known \cite{Lab,Bra69}  for type II superconductors close to the critical field
$H_{c2}$ \footnote{In order to receive Eq. (\ref{C-2}) from these papers one should use the
relation $gn^2= (H_{c2}-H)^2/8\pi \kappa^2 \beta$, which follows from the Ginzburg-Landau
theory in the limit $\kappa=\lambda/\xi \to \infty$.} and with the value obtained by Sinova {\em et al.} \cite{SHM} (after taking into account the
different definition of the elastic modulus $c_{66}$ by Sinova {\em et al}: $c_{66}=2C_{66}$). 

So in the LLL regime the Tkachenko velocity 
\be
c_T =\sqrt{C_{66} \over \rho}= 0.453 c_s
    \ee{c-T}
is of the same order as  the sound velocity, in contrast to the VLL regime where $c_T \sim c_s \xi/b$ is much smaller than $c_s$ because of small ratio $\xi/b$.

Equations (\ref{omegRap}) and (\ref{c-T}) yield a very simple expression for Tkachenko eigenfrequencies in the LLL regime:
\be
\omega_i= 0.641 \gamma_i  (\omega_\perp-\Omega).
    \ee{}
For the lowest eigenfrequency $i=1$ with $\gamma_1= 9.66$ [see the paragraph before \eq{omegRap}] this yields $\omega_1=6.19(\omega_\perp-\Omega)$. Note that the Tkachenko velocity in the LLL regime is smaller than its value $\sqrt{\kappa\Omega /8\pi}$ in the VLL regime because of small sound velocity $c_s$ in the LLL regime. However, the absolute values of the eigenfrequencies grow at the crossover from the VLL to the LLL regime because   the ratio $c_T/c_s$ grows at the crossover. 

Schweikhard {\em et al.} \cite{Corn4} increased the rotation speed in an attempt to reach the LLL
regime . They observed linear dependence of the Tkachenko eigenfrequency on small $\omega_\perp-\Omega$ as was predicted by the theory. On the basis of good quantitative agreement with the theoretical calculation for the LLL regime by Baym \cite{Ba} Schweikhard {\em et al.} concluded that they have already reached the LLL regime. 
However, the correct value of the shear modulus $C_{66}$ in Eq.~(\ref{C-2})
is 10 times larger than the value $(81/80 \pi^4) \rho  c_s^2$ obtained by Baym \cite{Ba} and used for comparison.  
The frequencies of the observed Tkachenko mode in fact about 4 times less than correct theoretical values for the LLL regime. It is evidence that
the  experiment has not yet reached the LLL limit. Since experimental
values of $(\omega_\perp-\Omega)/\omega_\perp$ look small enough, apparently in order to reach the LLL limit more closely, the experiment should be done with a smaller number of atoms.

Concluding this section, let us consider restrictions on the existence of the LLL regime. First, the energy of the
lowest Landau level, $\hbar \Omega$, should exceed the interaction energy per particle
$\beta gn \approx gn$. This yields the inequality $n \ll \hbar \Omega /g \sim n_v \hbar^2/mg  $.  Second, the BEC with a regular vortex lattice exists as far as the filling factor
$n/n_v$ (the number of particles per vortex) exceeds  unity (see below), i.e., the
inequality $n \gg \Omega/\kappa$ is required. The two inequalities determine the interval of filling factors, where the LLL regime exists:
\be
{\hbar^2\over mg} \gg {n\over n_v} \gg 1. 
       \ee{}
So the LLL regime is observable only  for  a
weakly interacting Bose gas when $g\ll h^2/m$. The latter inequality means that the coherence length $\xi \sim \kappa /c_s \sim \hbar /\sqrt{mgn}$ exceeds the interparticle distance $\sim 1/\sqrt{n}$. 

What should happen with the LLL regime when the filling factor $n/n_v$ approaches unity?
Possible answers to this question were investigated by theoreticians both numerically
and analytically \cite{SHM,Ba,CWG}. They expect melting of the vortex lattice and destruction of the Bose condensate. Naturally the Tkachenko mode would disappear in this case. 

\section{Tkachenko waves in a superfluid cold atom Fermi gas}

The theory of Tkachenko waves, both in an incompressible and in a compressible liquid, can be extended on Fermi superfluids \cite{WCS}. The difference between a Fermi and a Bose superfluid is in the equation of state, which connects the chemical potential and the particle density. However, numerical calculations by Watanabe {\em et al.} \cite{WCS}  of equations similar to Eqs.~(\ref{eq1}) and (\ref{eq2})  
have shown that the difference in the equation of state has a weak effect on the Tkachenko mode eigenfrequencies. 
At the same time Watanabe {\em et al.} pointed out that in a Fermi superfluid gas of cold atoms it is easier to provide a larger number of particles in the cloud. This can help to reach conditions when the effect of compressibility is weaker and it is easier to determine the circulation quantum from the Tkachenko eigenfrequencies. Tkachenko waves in a superfluid Fermi gas of cold atoms are still waiting their experimental observation.

\section{Conclusion}

The Tkachenko wave predicted a half a century ago remains an object of intensive theoretical and experimental investigations because they provide a valuable information on properties of the ordered vortex lattices. Now these investigations focus  on rotating cold atom superfluids. Here the first unambiguous observation of the Tkachenko mode was carried out.  Nowadays challenges for the experiments are observation of the Tkachenko wave  in the lowest Landau level regime  (mean-field quantum Hall regime) and in the Fermi superfluid gases. Since the existence of the Tkachenko mode is intimately connected with the crystalline order in the vortex array, the Tkachenko mode can be a probe of the vortex lattice melting at small filling factors. 

\end{document}